\documentclass[aps,prd,twocolumn,floatfix,nofootinbib]{revtex4-1}
\usepackage[dvipdfmx]{graphicx}
\usepackage{amsmath,amssymb,siunitx,url,color,hyperref}

\newcommand{\prx}{Phys. Rev. X}
\newcommand{\physrep}{Phys. Rep.}

\begin{document}
\title{Discrepancy in tidal deformability of GW170817 between the
Advanced LIGO twin detectors}
\author{Tatsuya Narikawa$^{1,2}$}
\author{Nami Uchikata$^{3,2}$}
\author{Kyohei Kawaguchi$^{2,4}$}
\author{Kenta Kiuchi$^{4,5}$}
\author{Koutarou Kyutoku$^{1,5,6,7}$}
\author{Masaru Shibata$^{4,5}$}
\author{Hideyuki Tagoshi$^2$}
\affiliation{$^1$Department of Physics, Kyoto University, Kyoto
606-8502, Japan}
\affiliation{$^2$Institute for Cosmic Ray Research, The University of
Tokyo, Chiba 277-8582, Japan}
\affiliation{$^3$Graduate School of Science and Technology, Niigata
University, Niigata 950-2181, Japan}
\affiliation{$^4$Max Planck Institute for Gravitational Physics (Albert
Einstein Institute), Am M{\"u}hlenberg 1, Potsdam-Golm 14476, Germany}
\affiliation{$^5$Center for Gravitational Physics, Yukawa Institute for
Theoretical Physics, Kyoto University, Kyoto 606-8502, Japan}
\affiliation{$^6$Theory Center, Institute of Particle and Nuclear
Studies, KEK, Tsukuba 305-0801, Japan}
\affiliation{$^7$Interdisciplinary Theoretical and Mathematical Sciences
Program (iTHEMS), RIKEN, Wako, Saitama 351-0198, Japan}
\date{\today}

\begin{abstract}
 We find that the Hanford and Livingston detectors of Advanced LIGO
 derive a distinct posterior probability distribution of binary tidal
 deformability $\tilde{\Lambda}$ of the first binary-neutron-star merger
 GW170817. By analyzing public data of GW170817 with a nested-sampling
 engine and the default TaylorF2 waveform provided by the LALInference
 package, the probability distribution of the binary tidal deformability
 derived by the LIGO-Virgo detector network turns out to be determined
 dominantly by the Hanford detector. Specifically, by imposing the flat
 prior on tidal deformability of individual stars, symmetric 90\%
 credible intervals of $\tilde{\Lambda}$ are estimated to be
 $527^{+619}_{-345}$ with the Hanford detector, $927^{+522}_{-619}$ with
 the Livingston detector, and $455^{+668}_{-281}$ with the LIGO-Virgo
 detector network. Furthermore, the distribution derived by the
 Livingston detector changes irregularly when we vary the maximum
 frequency of the data used in the analysis. This feature is not
 observed for the Hanford detector. While they are all consistent, the
 discrepancy and irregular behavior suggest that an in-depth study of
 noise properties might improve our understanding of GW170817 and future
 events.
\end{abstract}

\maketitle

\section{Introduction}

Tidal deformability of neutron stars can be a key quantity to understand
the hitherto-unknown nature of supranuclear density matter (see
Ref.~\cite{lattimer_prakash2016} for reviews). The relation between the
mass and tidal deformability is uniquely determined by the neutron-star
equation of state \cite{harada2001,lindblom_indik2014} as is the
mass--radius relation \cite{lindblom1992}. Thus, simultaneous
measurements of the mass and tidal deformability are eagerly desired,
and gravitational waves from binary-neutron-star mergers give us a
perfect opportunity. Once the mass--tidal deformability relation is
understood accurately, binary neutron stars can be used as standard
sirens to explore the expansion of the universe even in the absence of
electromagnetic counterparts \cite{messenger_read2012}. Motivated by
these facts, the influence of tidal deformability on gravitational waves
from binary neutron stars has been studied vigorously in this decade
\cite{flanagan_hinderer2008,damour_nagar2010,vines_fh2011,damour_nv2012,hotokezaka_ks2013,read_bcfgkmrst2013,hinderer_etal2016}.

The direct detection of gravitational waves from a binary-neutron-star
merger, GW170817, enabled us to measure the tidal deformability of a
neutron star for the first time \cite{ligovirgo2017-3}. The LIGO-Virgo
collaboration (LVC) reported an upper bound on the most influential
combination of tidal deformability parameters of two neutron stars, the
so-called binary tidal deformability $\tilde{\Lambda}$, to be $\lesssim
800$ (all the values in this paper refer to 90\% credibility) in their
discovery paper \cite{ligovirgo2017-3} under the reasonable assumption
of small neutron-star spins (later corrected to $\lesssim 900$
\cite{ligovirgo2019}). Independent analysis in Ref.~\cite{de_flbbb2018}
reported, e.g., $\tilde{\Lambda} = 222^{+420}_{-138}$ with the flat
prior on the mass of neutron stars and the reasonable assumption of a
common, causal equation of state for both neutron stars. LVC also
reported an updated highest-posterior-density interval, $\tilde{\Lambda}
= 300^{+420}_{-230}$ \cite{ligovirgo2019} using sophisticated waveform
models \cite{dietrich_bt2017,dietrich_etal2019} (see also
Ref.~\cite{ligovirgo2019-3} for an update), and this is further
restricted to $190^{+390}_{-120}$ if a common equation of state is
assumed \cite{ligovirgo2018}.

All these inferences are made by combining the output of Advanced LIGO
twin detectors, i.e., the Hanford and Livingston detectors (and Advanced
Virgo). It should be important to examine the extent to which results
derived by individual detectors agree, particularly in the presence of a
glitch near merger \cite{ligovirgo2017-3}. A study of
\textit{p}--\textit{g} instability presented the posterior probability
distribution of $\tilde{\Lambda}$ derived by individual detectors
\cite{ligovirgo2019-2}, but this is estimated only by incorporating this
effect and by allowing high spins, which broaden the distribution of
$\tilde{\Lambda}$. Neither consistency nor discrepancy of the derived
distribution is discussed.

In this paper, we present our independent analysis of GW170817 to show
that the Advanced LIGO twin detectors derive a distinct posterior
probability distribution of $\tilde{\Lambda}$ (and only for this
quantity; see the Appendix). Although the 90\% credible intervals of
$\tilde{\Lambda}$ are nominally consistent between the twin detectors,
the distribution derived by the Livingston detector tends to prefer
larger values of $\tilde{\Lambda}$ than those reported in the
literature. Close inspection of the distribution suggests that the
difference between the twin detectors might not be purely
statistical. Specifically, the distribution derived by the Livingston
detector does not behave smoothly with respect to the variation of the
maximum frequency of the data used for parameter estimation. This
behavior is not expected from physics of tidal deformation and should be
contrasted with that of the distribution derived by the Hanford
detector. The discrepancy between the twin detectors presages a
challenge for determining tidal deformability accurately in future
detections.

\section{Parameter estimation}

We perform a Bayesian parameter estimation of GW170817 for (1) the
Hanford data (Hanford-only), (2) the Livingston data (Livingston-only),
and (3) combined data of the twin detectors and Advanced Virgo (HLV) as
in previous work
\cite{ligovirgo2017-3,de_flbbb2018,ligovirgo2019-3,ligovirgo2019}. The
data of GW170817 are made public by
LVC.\footnote{\url{https://www.gw-openscience.org/catalog/GWTC-1-confident/single/GW170817/}
for Hanford and Virgo, \url{https://dcc.ligo.org/LIGO-T1700406/public}
for Livingston} The calibration error is taken into account by using the
calibration uncertainty envelope release for
GWTC-1.\footnote{\url{https://dcc.ligo.org/LIGO-P1900040/public}} As far
as we tested, results derived by the HLV data change very little when
the data from Virgo are discarded, as expected from the small
signal-to-noise ratio \cite{ligovirgo2017-3}.

Presuming that gravitational waves are detected in the relevant data,
$s(t)$, we compute the posterior probability distribution of binary
parameters $\boldsymbol{\theta}$ via
\begin{equation}
 p ( \boldsymbol{\theta} | s(t) ) \propto p ( s(t) | \boldsymbol{\theta}
  ) p ( \boldsymbol{\theta} ) ,
\end{equation}
where $p ( s(t) | \boldsymbol{\theta} )$ is the likelihood and $p (
\boldsymbol{\theta} )$ is the prior probability distribution. The
parameters, $\boldsymbol{\theta}$, consist of two masses, (aligned
components of) two spins, two tidal deformability parameters, the
luminosity distance, the sky position, the binary inclination, the
polarization angle, the coalescence time, and the coalescence phase
\cite{ligovirgo2017-3}.

We employ the nested sampling \cite{skilling2006,veitch_vecchio2010} for
the practical analysis using an engine implemented in the public
LALInference package \cite{veitch_etal2015}, a part of LSC Algorithmic
Library Suite. We checked that all the independent realizations of the
nested-sampling chains derive consistent results. We also checked that
Markov Chain Monte Carlo parameter estimation derive consistent results.

We evaluate the likelihood following the standard procedure (see, e.g.,
Refs.~\cite{jaranowski_krolak,creighton_andersson}) using the noise
power spectrum derived by the relevant data using BayesWave (see
Appendix A and Appendix B of
Ref.~\cite{ligovirgo2019-3}).\footnote{\url{https://dcc.ligo.org/LIGO-P1900011/public}}
The noise is assumed to be stationary and Gaussian during the parameter
estimation. Following previous work \cite{ligovirgo2017-3,de_flbbb2018},
we adopt the restricted post-Newtonian TaylorF2 approximant as the
waveform model (see Ref.~\cite{ligovirgo2019} and references
therein). This choice facilitates comparisons with previous
results. Because this approximant is implemented in LALInference,
results presented in this work should not be affected by our own
analysis method and should be easy to reproduce. The minimum frequency
of the data used in the analysis is fixed to \SI{23}{\hertz}, and the
maximum frequency $f_\mathrm{max}$ is varied to investigate its
influence on estimation of $\tilde{\Lambda}$. Because we truncate the
TaylorF2 approximant above the frequency at the innermost stable
circular orbit around a nonspinning black hole with its mass being equal
to the total mass of the binary, the results become identical for
$f_\mathrm{max} \gtrsim \SI{1600}{\hertz}$. In this work, we represent
them by $f_\mathrm{max} = \SI{2048}{\hertz}$ for simplicity.

We caution that the inferred value of $\tilde{\Lambda}$ entails
systematic errors associated with inaccuracy of the TaylorF2
approximant. While the systematic error is subdominant compared to the
statistical error for GW170817 \cite{ligovirgo2019}, we are also
conducting further analysis employing a sophisticated waveform model
developed based on numerical-relativity simulations by the Kyoto group
\cite{kiuchi_kksst2017,kawaguchi_kksst2018}. Regarding the topic of this
paper, preliminary results suggest that this model only enhances the
discrepancy between the twin detectors. In particular, the posterior
probability distribution derived by the HLV data begins to exhibit a
multiply-peaked structure consistently with the LVC analysis performed
employing other sophisticated waveform models
\cite{ligovirgo2019}. These results will be presented in a separate
publication focusing on the comparison among waveform models
\cite{narikawa_ukkkst}.

The prior probability distribution is chosen to follow those adopted in
the LVC analysis \cite{ligovirgo2019}, and we mention specific choices
made in this work. The sky position is fixed to the location determined
by optical followup observations \cite{utsumi_etal2017} to save the
computational cost. We checked that this has a negligible impact on
estimation of $\tilde{\Lambda}$. This is expected, because
$\tilde{\Lambda}$ is determined entirely by the phase of gravitational
waves, while the sky position affects only the amplitude (see also
Refs.~\cite{ligovirgoem2017-2,de_flbbb2018}). It should be cautioned
that the sky position cannot be determined to any accuracy by a single
detector \cite{fairhurst2009} in the absence of electromagnetic
information. The low-spin prior (see Ref.~\cite{ligovirgo2019}) is
adopted for the neutron-star spins for simplicity. The prior of the
tidal deformability is chosen to be flat in $[0:5000]$ for individual
components. This choice neglects the underlying equation of state, and
its appropriate incorporation will tighten the constraint on
$\tilde{\Lambda}$ \cite{de_flbbb2018,ligovirgo2018}. If we impose the
flat prior on $\tilde{\Lambda}$, the discrepancy between the twin
detectors is alleviated but remains. This alleviation is reasonable,
because the flat prior on $\tilde{\Lambda}$ gives weight to the
low-$\tilde{\Lambda}$ region where the discrepancy is mild as we see
below.

In this study, we focus primarily on the marginalized posterior
probability distribution of $\tilde{\Lambda}$. For completeness, we
present estimates of other parameters in the Appendix. The discrepancy
between the Advanced LIGO twin detectors is not observed significantly
for parameters other than binary tidal deformability.

\section{Posterior of tidal deformability}

\begin{figure}
 \includegraphics[width=.95\linewidth]{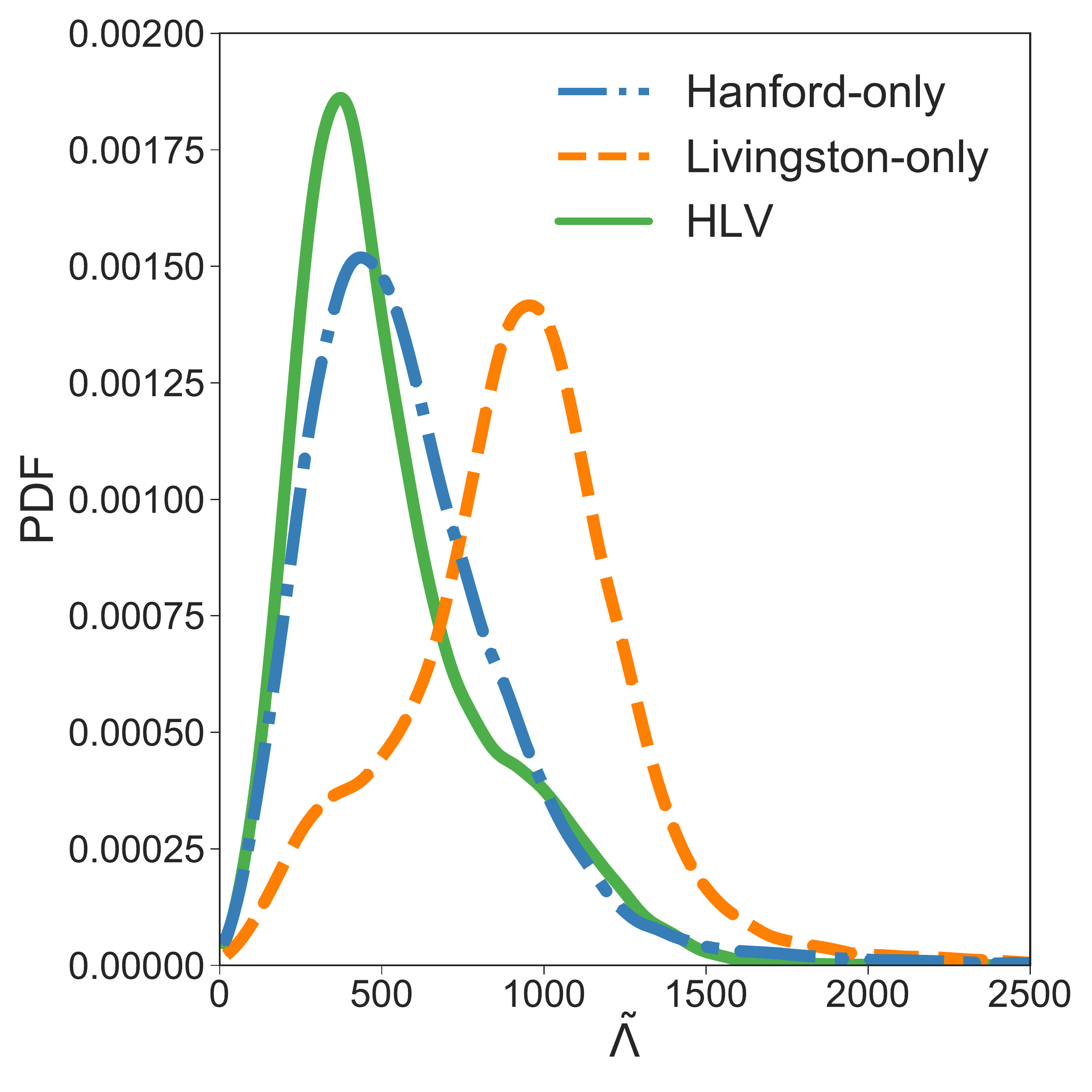} \caption{Marginalized
 posterior probability distribution of binary tidal deformability,
 $\tilde{\Lambda}$, derived by data of different detectors with
 $f_\mathrm{max} = \SI{2048}{\hertz}$. The symmetric 90\% credible
 intervals of $\tilde{\Lambda}$ for the Hanford-only data (blue) and the
 Livingston-only data (orange) are $\tilde{\Lambda} = 527^{+619}_{-345}$
 and $\tilde{\Lambda} = 927^{+522}_{-619}$, respectively. The
 distribution obtained by combined data of Advanced LIGO twin detectors
 and Advanced Virgo (HLV, green) is closer to that derived by the
 Hanford-only data with its symmetric 90\% credible interval being
 $\tilde{\Lambda} = 455^{+668}_{-281}$.} \label{fig:twin}
\end{figure}

\begin{table}
 \caption{90\% credible interval of binary tidal deformability,
 $\tilde{\Lambda}$, for different data and the maximum frequency,
 $f_\mathrm{max}$. The upper group shows the symmetric intervals, and
 the lower shows the highest-posterior-density intervals, where the
 median is shown as a representative value for both groups.}
 \begin{tabular}{cccc} \hline
  $f_\mathrm{max}$ & Hanford-only & Livingston-only & HLV \\
  \hline \hline
  \multicolumn{4}{c}{Symmetric interval} \vspace*{2pt} \\
  \SI{800}{\hertz} & $1109^{+860}_{-798}$ & $1023^{+750}_{-613}$ &
              $988^{+591}_{-556}$ \vspace{2pt} \\
  \SI{900}{\hertz} & $667^{+651}_{-461}$ & $913^{+729}_{-543}$ &
              $671^{+489}_{-400}$ \vspace{2pt} \\
  \SI{1000}{\hertz} & $598^{+613}_{-413}$ & $853^{+627}_{-479}$ &
              $660^{+461}_{-384}$ \vspace{2pt} \\
  \SI{1100}{\hertz} & $636^{+566}_{-423}$ & $785^{+601}_{-433}$ &
              $659^{+397}_{-351}$ \vspace{2pt} \\
  \SI{1200}{\hertz} & $573^{+564}_{-372}$ & $823^{+620}_{-449}$ &
              $594^{+500}_{-362}$ \vspace{2pt} \\
  \SI{1300}{\hertz} & $579^{+543}_{-372}$ & $889^{+573}_{-572}$ &
              $521^{+557}_{-288}$ \vspace{2pt} \\
  \SI{1400}{\hertz} & $540^{+598}_{-342}$ & $923^{+533}_{-611}$ &
              $480^{+657}_{-280}$ \vspace{2pt} \\
  \SI{1500}{\hertz} & $526^{+627}_{-325}$ & $925^{+526}_{-634}$ &
              $438^{+655}_{-253}$ \vspace{2pt} \\
  \SI{2048}{\hertz} & $527^{+619}_{-345}$ & $927^{+522}_{-619}$ &
              $455^{+668}_{-281}$ \vspace{2pt} \\
  \hline
  \multicolumn{4}{c}{Highest posterior density interval}
  \vspace*{2pt} \\
  \SI{800}{\hertz} & $1109^{+797}_{-857}$ & $1023^{+681}_{-677}$ &
              $988^{+574}_{-577}$ \vspace{2pt} \\
  \SI{900}{\hertz} & $667^{+556}_{-529}$ & $913^{+632}_{-620}$ &
              $671^{+451}_{-437}$ \vspace{2pt} \\
  \SI{1000}{\hertz} & $598^{+508}_{-479}$ & $853^{+549}_{-543}$ &
              $660^{+424}_{-417}$ \vspace{2pt} \\
  \SI{1100}{\hertz} & $636^{+484}_{-487}$ & $785^{+522}_{-495}$ &
              $659^{+378}_{-369}$ \vspace{2pt} \\
  \SI{1200}{\hertz} & $573^{+451}_{-443}$ & $823^{+530}_{-519}$ &
              $594^{+444}_{-409}$ \vspace{2pt} \\
  \SI{1300}{\hertz} & $579^{+438}_{-437}$ & $889^{+524}_{-615}$ &
              $521^{+422}_{-362}$ \vspace{2pt} \\
  \SI{1400}{\hertz} & $540^{+485}_{-410}$ & $923^{+468}_{-665}$ &
              $480^{+541}_{-351}$ \vspace{2pt} \\
  \SI{1500}{\hertz} & $526^{+505}_{-399}$ & $925^{+467}_{-684}$ &
              $438^{+546}_{-320}$ \vspace{2pt} \\
  \SI{2048}{\hertz} & $527^{+498}_{-419}$ & $927^{+467}_{-666}$ &
              $455^{+562}_{-349}$ \vspace{2pt} \\
  \hline
 \end{tabular}
 \label{table:tidal}
\end{table}

Figure \ref{fig:twin} shows the marginalized posterior probability
distribution of binary tidal deformability, $\tilde{\Lambda}$, derived
by the Hanford-only data, Livingston-only data, and combined HLV data
with $f_\mathrm{max} = \SI{2048}{\hertz}$. The corresponding 90\%
credible intervals are presented in Table \ref{table:tidal}. The HLV
distribution exhibits a peak at $\tilde{\Lambda} \approx 370$ with a
tail extending to the high-$\tilde{\Lambda}$ region consistently with
previous work \cite{de_flbbb2018,ligovirgo2019}. Estimates of other
parameters are also broadly consistent with those derived in previous
work (see the Appendix).

The separate analysis of the data obtained by the individual of twin
detectors reveals a mild discrepancy between them. On the one hand, the
posterior probability distribution derived by the Hanford-only data is
similar to that derived by the HLV data. It exhibits a peak at
$\tilde{\Lambda} \approx 440$ with a tail at the high-$\tilde{\Lambda}$
region. Because the tail structure is enhanced for the HLV data due to
the Livingston detector as we discuss below, the 90\% credible intervals
are similar for these two distributions (see Table
\ref{table:tidal}). On the other hand, the distribution derived by the
Livingston-only data peaks at a large value of $\tilde{\Lambda} \approx
960$, which is close to the edge of the 90\% credible intervals of
either the Hanford-only or HLV distribution. Conversely, the
Livingston-only distribution has a small probability around
$\tilde{\Lambda} \approx 400$.

We find that the HLV distribution is approximately reproduced by
multiplying the Hanford-only distribution and the Livingston-only
distribution if we appropriately incorporate the prior probability
distribution of $\tilde{\Lambda}$ determined by that of other
parameters. Specifically, we need to divide the multiplied distribution
by the prior, because it is included in both the Hanford-only and
Livingston-only distributions. The division by the prior reduces the
probability at the high-$\tilde{\Lambda}$ region, and thus the data of
the Livingston detector have a minor influence on the combined
result. Still, the HLV distribution shows a bump at $\tilde{\Lambda}
\approx 1000$ inherited from the peak of the Livingston-only
distribution.

\begin{figure}
 \includegraphics[width=.95\linewidth]{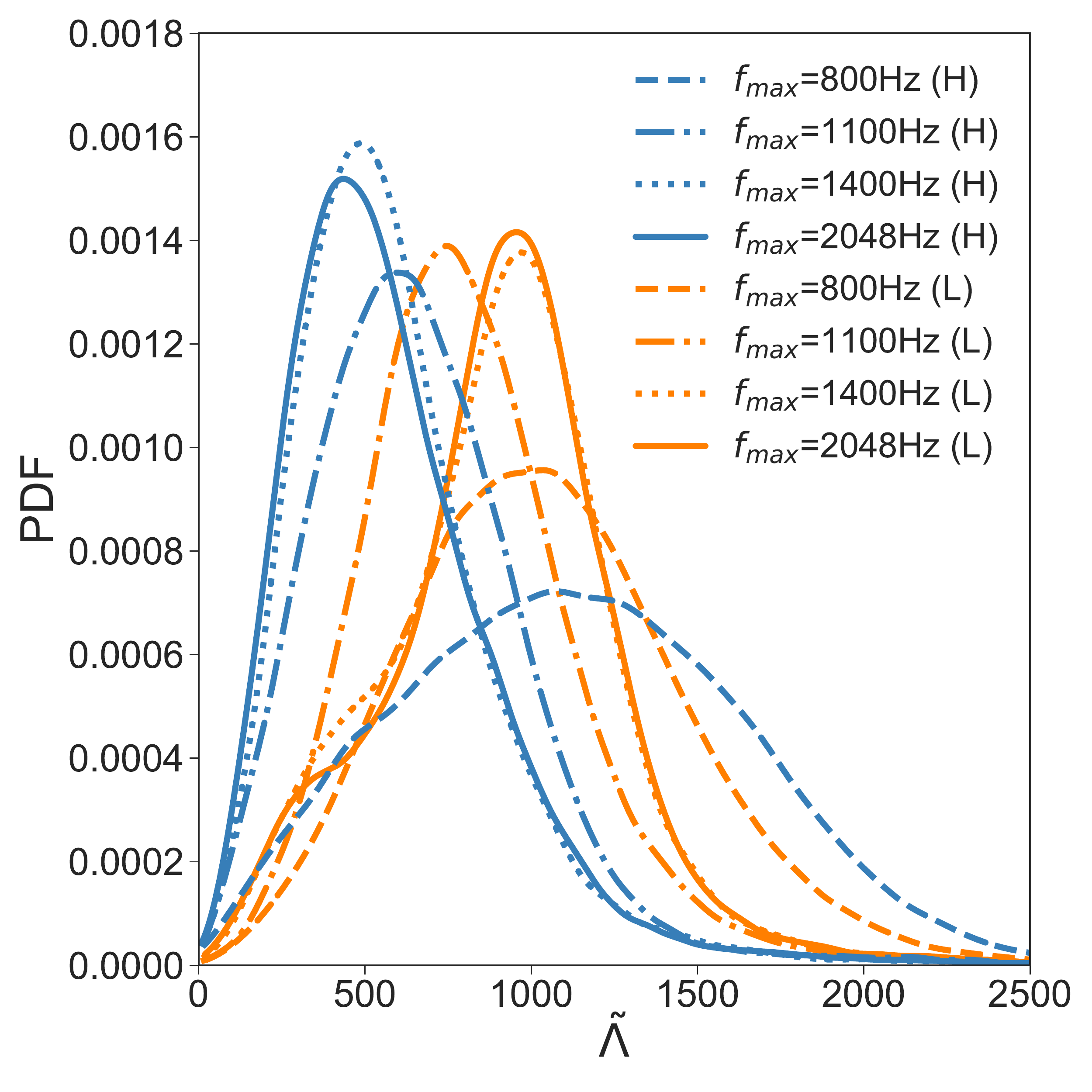} \caption{Dependence of
 the marginalized posterior probability distribution on the maximum
 frequency, $f_\mathrm{max}$, imposed in the data analysis. We adopt
 $f_\mathrm{max} = 800$, 1100, 1400, and \SI{2048}{\hertz}. While the
 distribution derived by the Hanford-only data (blue, denoted by H)
 shifts smoothly to the low-$\tilde{\Lambda}$ region as $f_\mathrm{max}$
 increases, that by the Livingston-only data (orange, denoted by L)
 shows a turn at $f_\mathrm{max} \approx \SI{1100}{\hertz}$.}
 \label{fig:fmax}
\end{figure}

Detailed features of the data from individual detectors are clarified by
examining changes of the posterior probability distribution with respect
to the variation of the maximum frequency, $f_\mathrm{max}$, imposed in
the data analysis. Figure \ref{fig:fmax} shows the results for
$f_\mathrm{max}=800$, 1100, 1400, and \SI{2048}{\hertz} (the same as
Fig.~\ref{fig:twin}). The Hanford-only distribution shifts smoothly to
the low-$\tilde{\Lambda}$ region as $f_\mathrm{max}$ increases, and the
median value decreases approximately monotonically. This is reasonably
expected, because the tidal deformability is primarily determined by the
gravitational-wave data at high frequency
\cite{damour_nv2012,de_flbbb2018}. This feature results from the nature
of tidal interaction and does not rely on complicated relativistic
effects.

On another front, the posterior probability distribution derived by the
Livingston-only data shows irregular behavior with respect to the
increase of $f_\mathrm{max}$. At the beginning, the distribution shifts
to the low-$\tilde{\Lambda}$ region in a similar manner to the
Hanford-only distribution when $f_\mathrm{max}$ is decreased from
\SI{800}{\hertz}. However, the shift turns around at $f_\mathrm{max}
\approx \SI{1100}{\hertz}$, and then the distribution moves back to the
high-$\tilde{\Lambda}$ region. The median values presented in Table
\ref{table:tidal} clearly exhibit the turn, and neither a systematic
decrease nor increase is observed. This is not naturally anticipated
from the viewpoint of measurability \cite{damour_nv2012,de_flbbb2018}.

Taking the fact that the distribution of $\tilde{\Lambda}$ obtained by
the combined HLV data is similar to that by the Hanford-only data into
account, the peculiar dependence on $f_\mathrm{max}$ might indicate that
the high-frequency data of the Livingston detector are not very helpful
to determine $\tilde{\Lambda}$ of GW170817. Note that the
signal-to-noise ratio is larger for Livingston than for Hanford due to
the higher sensitivity \cite{ligovirgo2017-3,ligovirgo2019}.
Specifically, the signal-to-noise ratios derived in our analysis are
18.8, 26.8, and 32.6 for Hanford, Livingston, and HLV,
respectively. Thus, the dominance of Hanford data in the combined result
is not reasonably understood from the signal-to-noise ratio. Actually,
we find that the posterior probability distributions of other parameters
are more strongly influenced by Livingston than Hanford (see the
Appendix).

A careful examination of Fig.~\ref{fig:fmax} reveals that a small bump
appears for $f_\mathrm{max} \gtrsim \SI{1400}{\hertz}$ at the
low-$\tilde{\Lambda}$ region of the posterior probability distribution
derived by the Livingston-only data. The location of this bump is close
to the peak of $\tilde{\Lambda}$ derived by the Hanford-only data. Thus,
the Livingston-only distribution may consist of the main peak with a
large value of $\tilde{\Lambda}$ associated with the low-frequency data
and the side peak with a small value of $\tilde{\Lambda}$ associated
with the high-frequency data.

\section{Discussion}

Our analysis suggests that the noise in the high-frequency region of the
Livingston data somehow corrupted information about the tidal
deformability of GW170817. Although this could simply be caused by the
stationary and Gaussian noise, it could be worthwhile to look for
possible peculiar features in the data of GW170817, e.g., a residual of
the glitch in the Livingston data at about a half second before merger
\cite{ligovirgo2017-3} (but see also Ref.~\cite{pankow_etal2018}) as it
is quite important to estimate tidal deformability accurately. By
contrast, the Hanford data seem to be well-behaved during the reception
of GW170817.

Having said that, it is ultimately impossible to judge which of the
Advanced LIGO twin detectors provides us with a more reliable estimate
of the binary tidal deformability than the other does, or whether simply
their combination is the most reliable, without meaningful data obtained
by a third detector. It should be emphasized that the 90\% credible
intervals are consistent between the twin detectors. What we may safely
conclude is that the posterior probability distribution is exceptionally
distinct for binary tidal deformability (see the Appendix for other
parameters) and that the Livingston data are not very useful for
constraining its value in the case of GW170817. Secure parameter
estimation will be helped by unambiguous detection by other instruments
such as Advanced Virgo or KAGRA \cite{ligovirgo2016-lr}. However, if the
irregular loss of information is typical for detections with a moderate
signal-to-noise ratio, accurate determination of tidal deformability
will remain challenging unless its origin is identified.

\begin{acknowledgments}
 We thank John Veitch for a very helpful explanation of LALInference and
 Chris van den Broeck for useful discussions. We also thank Soichiro
 Morisaki, extreme-matter co-chairs, ROTA members for GW170817, and
 members of the Gravitational Wave research group at Nikhef for
 discussions on the noise power spectrum. This work is supported by
 Japanese Society for the Promotion of Science (JSPS) KAKENHI Grants
 No.~JP15K05081, JP16H02183, JP16H06342, JP17H01131, JP17H06358,
 JP17H06361, JP18H01213, JP18H04595, and JP18H05236, and by a post-K
 project hp180179. This work is also supported by JSPS Core-to-Core
 Program A, Advanced Research Networks and by the joint research program
 of the Institute for Cosmic Ray Research, University of Tokyo, and the
 Computing Infrastructure Project of KISTI-GSDC in the Republic of
 Korea. T.~Narikawa was supported in part by a Grant-in-Aid for JSPS
 Research Fellows, and he is also grateful to the hospitality of Chris's
 group during his stay at Nikhef. K.~Kawaguchi was supported in part by
 JSPS overseas research fellowships. We are also grateful to the
 LIGO-Virgo collaboration for the public release of gravitational-wave
 data of GW170817. This research has made use of data, software, and web
 tools obtained from the Gravitational Wave Open Science Center
 (\url{https://www.gw-openscience.org}), a service of LIGO Laboratory,
 the LIGO Scientific Collaboration, and the Virgo Collaboration. LIGO is
 funded by the U.~S. National Science Foundation. Virgo is funded by the
 French Centre National de la Recherche Scientifique (CNRS), the Italian
 Istituto Nazionale di Fisica Nucleare (INFN), and the Dutch Nikhef,
 with contributions by Polish and Hungarian institutes.
\end{acknowledgments}

\appendix

\section{Parameter other than tidal deformability} \label{app:other}

\begin{table*}
 \caption{90\% credible interval of the luminosity distance, the binary
 inclination, mass parameters, and the effective spin parameter derived
 by different data with $f_\mathrm{max} = \SI{2048}{\hertz}$. We show
 10\%--100\% regions of the mass ratio with the upper limit $q=1$
 imposed by the prior, and those of $m_1$ and $m_2$ are given
 accordingly. We give symmetric 90\% credible intervals, i.e.,
 5\%--95\%, for the other parameters with the median as a representative
 value. The binary inclination derived by either the Hanford-only or
 Livingston-only data is undetermined up to the orbital-plane
 reflection, and thus the 90\% credible interval is very large.}
 \begin{tabular}{lccc} \hline
  & Hanford-only & Livingston-only & HLV \\
  \hline \hline
  Luminosity distance $d_L~[\si{Mpc}]$ & $40.8^{+10.3}_{-17.1}$ &
          $39.0^{+9.0}_{-16.4}$ & $41.0^{+7.7}_{-15.2}$ \vspace{2pt} \\
  Binary inclination $\theta_\mathrm{JN} $ [degree] & $88^{+77}_{-74}$ &
          $86^{+79}_{-72}$ & $146^{+24}_{-28}$ \vspace{2pt} \\
  Chirp mass in the detector frame $\mathcal{M}^\mathrm{det}~[M_\odot]$
  & $1.1978^{+0.0002}_{-0.0002}$ & $1.1975^{+0.0001}_{-0.0001}$ &
              $1.1975^{+0.0001}_{-0.0001}$ \vspace{2pt} \\
  Chirp mass in the source frame $\mathcal{M}~[M_\odot]$ &
      $1.187^{+0.005}_{-0.003}$ & $1.187^{+0.004}_{-0.002}$ &
              $1.187^{+0.004}_{-0.002}$ \vspace{2pt} \\
  Primary mass $m_1~[M_\odot]$ & (1.36, 1.56) & (1.36, 1.59) & (1.36,
              1.59) \vspace{2pt} \\
  Secondary mass $m_2~[M_\odot]$ & (1.20, 1.37) & (1.17, 1.37) & (1.17,
              1.37) \vspace{2pt} \\
  Total mass $M_\mathrm{tot}:=m_1+m_2~[M_\odot]$ &
      $2.73^{+0.03}_{-0.01}$ & $2.74^{+0.04}_{-0.01}$ &
              $2.74^{+0.04}_{-0.01}$ \vspace{2pt} \\
  Mass ratio $q:=m_2/m_1$ & (0.77, 1) & (0.74, 1) & (0.74, 1)
              \vspace{2pt} \\
  Effective spin $\chi_\mathrm{eff}$ & $0.014^{+0.010}_{-0.011}$ &
          $0.003^{+0.015}_{-0.010}$ & $0.002^{+0.015}_{-0.009}$
              \vspace{2pt} \\
  \hline
 \end{tabular}
 \label{table:other}
\end{table*}

\begin{figure*}
 \includegraphics[width=1.\linewidth]{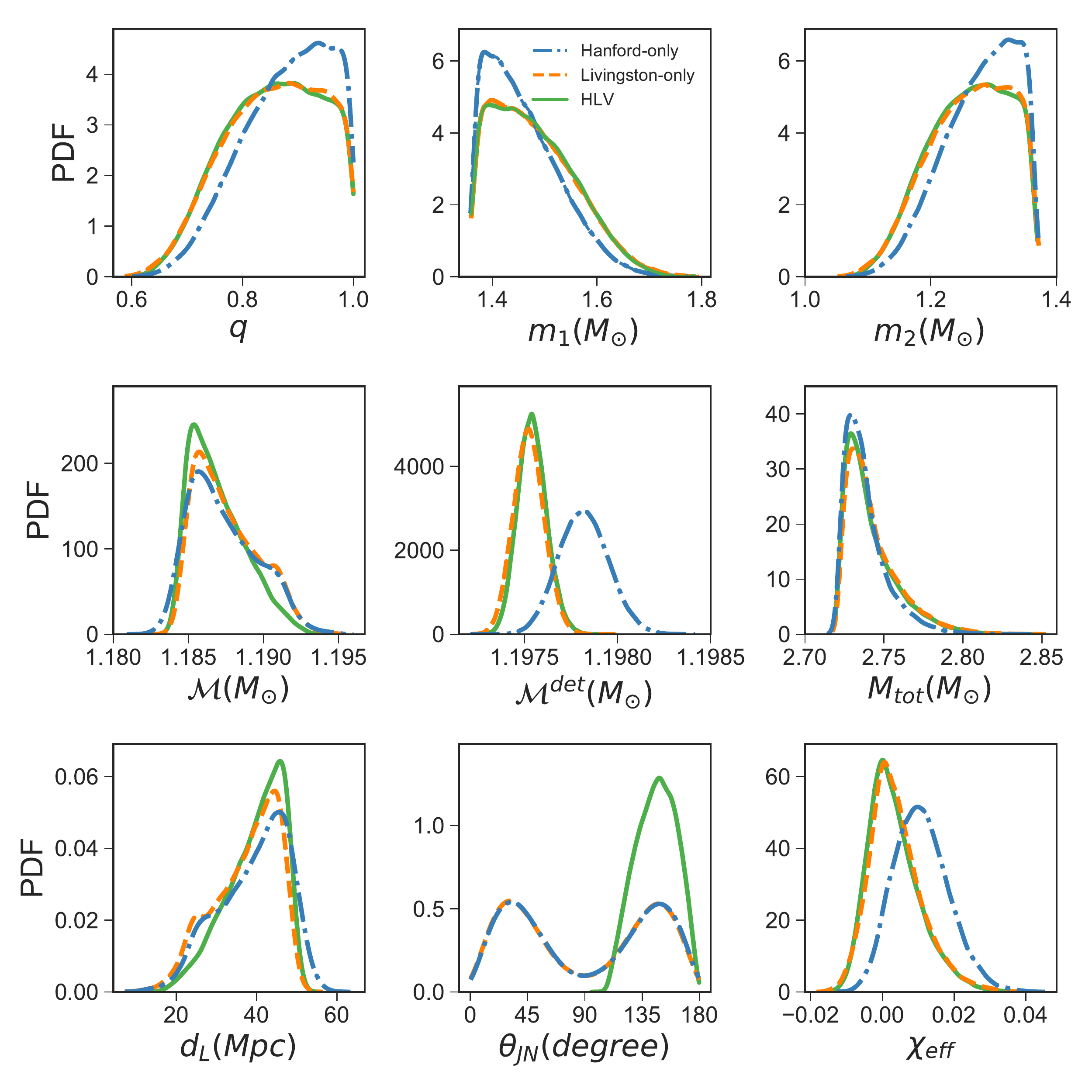} \caption{Marginalized
 posterior probability distribution of various parameters with
 $f_\mathrm{max} = \SI{2048}{\hertz}$. The blue, orange, and green
 curves correspond to the Hanford-only, Livingston-only, and HLV data,
 respectively. The top-left, top-middle, top-right, middle-left, center,
 middle-right, bottom-left, bottom-middle, and bottom-right panels show
 $q$, $m_1$, $m_2$, $\mathcal{M}$, $\mathcal{M}^\mathrm{det}$,
 $M_\mathrm{tot}$, $d_L$, $\theta_\mathrm{JN}$, and $\chi_\mathrm{eff}$,
 respectively (see Table \ref{table:other} for the definition of
 quantities). The distribution of $\theta_\mathrm{JN}$ for a
 single-detector data exhibits a bimodal structure due to the degeneracy
 of the orbital-plane reflection, and this is resolved for the HLV
 data.}  \label{fig:other}
\end{figure*}

We present estimates of parameters other than binary tidal deformability
with $f_\mathrm{max}=\SI{2048}{\hertz}$ for completeness. Table
\ref{table:other} presents the 90\% credible intervals of the luminosity
distance, the binary inclination, mass parameters, and the effective
spin parameter derived by different data. We recall that the sky
location is fixed by the information from electromagnetic observations
and that the low-spin prior is imposed. The cosmological redshift is not
taken from the host galaxy NGC4993 \cite{ligovirgoem2017-2} and is
determined from the luminosity distance by assuming the Hubble constant
$H_0 = \SI{69}{km.s^{-1}.Mpc^{-1}}$ (a default value in LAL) to derive
the chirp mass in the source frame. The marginalized posterior
distribution is presented in Fig.~\ref{fig:other}. The consistency of
our results with previous work \cite{de_flbbb2018,ligovirgo2019} serves
as an important sanity check. They also clarify that the estimation of
binary tidal deformability is exceptionally delicate for GW170817.

Table \ref{table:other} shows that the credible intervals agree
remarkably between the Hanford-only and Livingston-only data. The
marginalized posterior probability distribution depicted in
Fig.~\ref{fig:other} not only confirms this agreement but also shows
that the distribution is approximately identical. The parameters shown
here are estimated primarily from information at low frequency, where
the gravitational-wave signal spends most of time with a large number of
cycles \cite{damour_nv2012,de_flbbb2018}. The detector noise is also
less severe at lower frequency. As a result, the situation is different
from that of binary tidal deformability discussed in the main text.

The 90\% credible intervals derived by combining the HLV data are very
close to those of a single detector (Hanford-only or Livingston-only)
except for the binary inclination, $\theta_\mathrm{JN}$. The reason for
the difference in $\theta_\mathrm{JN}$ is that the degeneracy of the
reflection with respect to the orbital plane, i.e., face-on or face-off,
is resolved when the HLV data are combined. The resolution is clearly
shown in the middle panel of the bottom row of Fig.~\ref{fig:other},
where the bimodal structure for a single detector is changed to a single
peak for the HLV data favoring the face-off orientation. The posterior
probability distribution of other parameters shifts only moderately.

Close inspection reveals that the posterior probability distribution
derived by the HLV data closely follows that of the Livingston-only data
for the chirp mass in the detector frame, the mass ratio, and the
effective spin. Because they are determined by the gravitational-wave
phase at low frequency, the closeness of the distribution can be
understood as a result of the large signal-to-noise ratio for the
Livingston detector. Again, these features are different from what we
observe for the binary tidal deformability, where the distribution
derived by the HLV data is closer to that derived by the Hanford-only
data than the Livingston-only data.

%\bibliography{paper}
%

\end{document}